\documentclass[nofootinbib,aps,11pt]{revtex4}
\usepackage{graphicx}
\usepackage{cancel}
\usepackage{amssymb}
\usepackage{textcomp}
\usepackage{amsmath}
\usepackage{bm}
\usepackage{times}
\usepackage{hyperref}
\usepackage{epsfig}
\usepackage{color}
\usepackage{axodraw}

\begin{document}

\title{\Large 
Non-renormalizable Operators for Solar Mass Generation in
Split SuSy with Bilinear R-parity Violation.
                   }
\author{Marco Aurelio D\'\i az${}^1$}
\author{Benjamin Koch${}^1$}
\author{Nicol\'as Rojas${}^2$}
\affiliation{
{\small ${}^1$Instituto de F\'\i sica, Pontificia Universidad Cat\'olica 
de Chile, Avenida Vicu\~na Mackenna 4860, Santiago, Chile \\
\small ${}^2$Instituto de F\'isica Corpuscular CSIC/Universitat de Valencia, Parc Cient\'ific,
calle Catedr\'atico Jos\'e Beltr\'an, 2, E-46980 Paterna, Spain.
}}
%%%%%%%%%%%%%%%%%%%%%%%%%%%%%%%%%%%%%%%%%%%%%%%%%%%%%%%%%%%%%%%%%%%%
\begin{abstract}

The Minimal Supersymmetric Extension of the Standard Model (MSSM)
is able to explain the current data from neutrino physics.
Unfortunately Split Supersymmetry as low energy approximation of this theory
fails to generate a solar square mass difference,
including after the addition of bilinear R-Parity Violation.
In this work, it is shown how one can derive
an effective low energy theory from the MSSM
in the spirit of Split Supersymmetry, which has the potential
of explaining the neutrino phenomenology.
This is achieved by going beyond leading order in the process
of integrating out heavy scalars from the original theory, which results
in non-renormalizable operators in the effective low energy theory.
It is found that in particular a $d=8$ operator is crucial for 
the generation of the neutrino mass differences.

\end{abstract}
%%%%%%%%%%%%%%%%%%%%%%%%%%%%%%%%%%%%%%%%%%%%%%%%%%%%%%%%%%%%%%%%%%%%
\date{\today}
\maketitle
%\tableofcontents

%%%%%%%%%%%%%%%%%%%%%%%%%%%%%%%%%%%%%%%%%%%%%%%%%%%%%%%%%%%%%%%%%%%%
\section{Introduction}
%%%%%%%%%%%%%%%%%%%%%%%%%%%%%%%%%%%%%%%%%%%%%%%%%%%%%%%%%%%%%%%%%%%

{}The Standard Model (SM) of particle physics provides a theoretical explanation
for the experimental evidence on weak interactions and the masses of the
corresponding mediators. This last point has its climax with the discovery
of the Higgs boson in 2012 \cite{Aad:2012tfa,Chatrchyan:2012xdj}. 
Even though these evidences have 
been important in the historical developing of the SM, the theory is far 
from complete since neutrino masses and Dark Matter appear to be issues 
that cannot be explained by the original 
theory.\\

{}The proposal of Weinberg's $d=5$ operator \cite{Weinberg:1979sa} points 
towards a way to extend the SM in order to include an explanation
for neutrino masses, 
such that realizations of this operator at tree level 
and one loop can be achieved.
Other proposals can be found in \cite{Zee:1985id,Aoki:2008av}.
In this context, in order to obtain Majorana neutrino masses it is necessary
to have some mechanism to provide lepton number violation as well. 
When exploring supersymmetric extensions of the SM, 
lepton number and baryon number can be 
incorporated in an {\it ad hoc} symmetry 
called {\it R-Parity}. Amongst the proposals in this area, we are 
interested exclusively in the addition of bilinear R-parity violation to the minimal
supersymmetric extension of the SM (MSSM) as a way to study neutrino
physics.\\

{}On top of that, within the working hypothesis of a supersymmetric extension 
of the SM, at the present time data from ATLAS and CMS have no direct evidence 
for the observation of {\it sparticles} \cite{Aad:2015baa}. 
This seems to indicate that the supersymmetric scalar particles may be too massive 
to be produced at the LHC. In this context it is appealing to study models 
like Split SuSy (SS) 
\cite{ArkaniHamed:2004fb,Giudice:2004tc}, where the claim 
is that the sfermions and one Higgs doublet are very heavy and therefore they are decoupled 
from the low-energy theory. In the original proposal, sfermions are pushed 
to live at scales in the order of $10^{10}$ GeV where the testing of 
sfermions at present colliders turns out to be very difficult.
Moreover, it leads to an {\it unnatural} model since corrections to Higgs mass
must be very fine tuned in order to give rise to a Higgs of $125$ GeV. 
However, such problems can be avoided for Split SuSy if the sfermion masses
lie around $10^4$ GeV. Such a scale allows for a Higgs mass determined by
experiments, and thus having the possibility to test effects of heavy
sfermions in the present time experiments\cite{Giudice:2004tc,Binger:2004nn}.\\

{}The MSSM plus BRpV is able to generate two neutrino mass 
eigenvalues at loop level. If we combine Split SuSy and BRpV in one single
model, though, the decoupling of scalars retaining only the 
operators with lowest mass dimension (renormalizable), leaves a theory that 
cannot explain the observed neutrino oscillations, since it cannot produce a 
solar neutrino mass scale. Thus, one faces a
problem, on the one hand it seems that Split SuSy is favored from collider phenomenology, 
while on the other hand it seems to be disfavored by neutrino oscillation 
data \cite{Diaz:2006ee}.
One alternative to solve this tension is to 
include new physics in terms of gravity motivated operators in order to achieve 
neutrino masses \cite{Barbieri:1979hc,Diaz:2009yz}.
Nevertheless, it would be much more attractive to solve this problem 
within Split SuSy
as a low energy approximation to the MSSM.\\

%%{}Although other supersymmetric mechanisms can be invoked in order to
%%generate neutrino masses, we also want to highlight that the inclusion
%%of higher dimensional operators has been discussed in the context of
%%supersymmetry and BRpV although neither in this context nor in the operators
%%we will include.\\

{}The origin of this tension between collider and neutrino phenomenology
within the simplest version of Split SuSy can be understood in the light 
of the underlying supersymmetric theory. A quite general theorem states 
that within this theory the existence of a tree level neutrino Majorana 
mass implies the appearance of a B-L violating superpartner {\it Majorana}-like 
mass term and vice versa \cite{Hirsch:1997vz}. This is exactly what 
happens in the MSSM with RpV terms. Thus, within conventional Split SuSy it is actually 
not possible to accommodate both neutrino and collider phenomenology. The 
problem seems to be that after integrating out the scalars and retaining
the resulting renormalizable operators only, the interplay between the 
neutrino Majorana mass and the superpartner scalars is lost, and with it, 
the generation of a second non-zero neutrino mass.\\

{}The aim of this paper is to investigate the possibility
to accommodate neutrino oscillation physics within Split SuSy 
in a self consistent way (i.e. with no addition of operators motivated 
from physics outside the MSSM as a UV complete theory) when going beyond 
the leading term after integrating out the scalar fields. Other alternatives within this
context have been investigated (see for instance \cite{Chun:2004mu} and 
\cite{Dhuria:2011ye}), though, we will stick ourselves to original proposal
of Split SuSy and BRpV. 
Clearly it would be favorable to 
be able to explain the solar mass difference
without the incorporation of new physics and new couplings.
In this spirit it is shown that already retaining next to leading operators 
from the original MSSM Lagrangian gives rise to non-renormalizable higher 
dimensional operators in the effective low energy theory, which can reconcile 
the Split SuSy with neutrino oscillation data.

%%%%%%%%%%%%%%%%%%%%%%%%%%%%%%%%%%%%%%%%%%%%%%%%%%%%%%%%%%%%%%%%%%
\section{Split Supersymmetry from minimal integration of scalars}
%%%%%%%%%%%%%%%%%%%%%%%%%%%%%%%%%%%%%%%%%%%%%%%%%%%%%%%%%%%%%%%%%%

In Split Supersymmetry \cite{ArkaniHamed:2004fb} all sfermions together 
with one heavy Higgs doublet have a large mass, which is taken for simplicity 
to be degenerate at the scale $\widetilde m$. 

In this framework, since the
effects of sfermions appear at very large scale, the hierarchy problem is 
not solved. It is argued that gauge unification is still achieved, and a 
Dark Matter candidate is still present (the neutralino) when R-Parity is 
conserved. In addition, large flavour and CP violation are avoided. This 
type of model is supported by the fact the LHC collaborations do not see 
sfermions up to a mass near the TeV scale~\cite{Aad:2015baa}.
Split SuSy can be obtained from the MSSM which includes two Higgs doublets 
$H_u$ and $H_d$ \cite{Gunion:1984yn,Gunion:1989we}. The Higgs 
potential in the MSSM is
\begin{eqnarray}
V &=& \left( m_{H_u}^2+|\mu|^2 \right) H_u^\dagger H_u + \left( m_{H_d}^2+|\mu|^2 \right) H_d^\dagger H_d -
B_\mu \left( H_u^T i\sigma_2 H_d+h.c. \right) +
\nonumber \\ &&
\frac{1}{8} (g^2+g'^2) \left( H_u^\dagger H_u-H_d^\dagger H_d \right)^2 +
\frac{1}{2} g^2 |H_u^\dagger H_d|^2,
\label{HiggsPotMSSM}
\end{eqnarray}
where $m_{H_u}^2$ and $m_{H_d}^2$ are soft mass parameters corresponding to the $H_u$ and $H_d$ bosons,
$\mu$ is the supersymmetric Higgs mass parameter, and $B_\mu$ is the soft Higgs mixing
parameter (with units of mass square).

It is assumed that one Higgs doublet, $H_{SM}$, remains light and resembles the SM Higgs doublet.
The second one, $H_{SS}$, is heavy and for simplicity its 
components have a degenerate mass equal to 
$\widetilde m$. This doublet comes from a mixing of $H_u$ and $H_d$, given by
\begin{equation}
\left[ \begin{array}{c} H_{SM} \cr i\sigma_2 H_{SS}^* \end{array} \right] =
\left[ \begin{array}{cc} \cos\alpha & -\sin\alpha \cr \sin\alpha & \cos\alpha \end{array} \right]
\left[ \begin{array}{c} H_u \cr i\sigma_2 H_d^* \end{array} \right],
\end{equation}
where the mixing angle $\alpha$ is reminiscent of the MSSM. In the decoupling limit of the MSSM 
(equivalent to the Split Supersymmetric case) we have $\sin\alpha=-\cos\beta$ and 
$\cos\alpha=\sin\beta$, under the usual conventions for the angles in the MSSM \cite{Gunion:1989we}. 
In SS, the angle $\alpha=\beta-\pi/2$ is such that there is no mixing between 
$H_{SM}$ and $H_{SS}$. The mass terms for the two doublets are $m_h\approx 125$ GeV and 
$\widetilde m$ respectively.

The transition from the MSSM to Split SuSy is achieved by integrating out
the heavy scalars by solving their equation of motion. 
For instance, the heavy Higgs field can be replaced by
\begin{eqnarray}\label{HSS1}
H_{SS} \approx -\frac{1}{\widetilde m^2} \frac{c_\alpha}{\sqrt{2}}
\left[ g(\sigma\cdot\widetilde W)-g'\widetilde B \right] \widetilde H_d +
\frac{1}{\widetilde m^2} \frac{s_\alpha}{\sqrt{2}} i\sigma_2
\left[ g(\sigma^*\cdot\overline{\widetilde W})+g'\overline{\widetilde B} \right] \overline{\widetilde H_u}  + ...
\end{eqnarray}
In the original proposal of Split SuSy this scalar was also taken to be 
heavy, therefore the limit $\tilde{m} \rightarrow \infty$ was invoked, which 
decouples completely the field $H_{SS}$. After replacing (\ref{HSS1}) in
the Lagrangian of the MSSM and after decoupling the heavy Higgs, the R-Parity conserving part of the 
Split Supersymmetric Lagrangian reads~\cite{ArkaniHamed:2004fb,Giudice:2004tc,Binger:2004nn}
\begin{eqnarray}
{\cal L}^{split}_{RpC} &=& {\cal L}^{split}_{kinetic} \ + \ m^2H^\dagger 
H - \frac{\lambda}{2}(H^\dagger H)^2 
-\Big[ Y_u \overline q_L u_R i \sigma_2 H^* \ + \ Y_d \overline q_L d_R H \ 
+ \ Y_e \overline l_L e_R H \ + \
\nonumber\\ 
&&
+ \frac{M_3}{2} \widetilde G\widetilde G \ + \ \frac{M_2}{2} \widetilde W 
\widetilde 
W \ + \ \frac{M_1}{2} \widetilde B \widetilde B \ + \ 
\mu \widetilde H_u^T i \sigma_2 \widetilde H_d \ + \
\label{LagSplit}\\ &&
+\textstyle{\frac{1}{\sqrt{2}}} H^\dagger
(\tilde g_u \sigma\widetilde W \ + \ \tilde g'_u\widetilde B)\widetilde H_u
 \ + \ \textstyle{\frac{1}{\sqrt{2}}} H^T i \sigma_2
(-\tilde g_d \sigma \widetilde W+ \tilde g'_d \widetilde B)\widetilde H_d 
+\mathrm{h.c.}\Big],
\nonumber
\end{eqnarray}
where for simplicity we have called $H$ the SM Higgs doublet. This Lagrangian contains
 the mass and quartic parameters of the light Higgs potential, 
the Yukawa couplings, the gaugino mass parameters, the supersymetric higgsino mass
parameter, and four Higgs-gaugino-higgsino couplings $\tilde g$, by-products of the
decoupling of the scalars.

The original Split Supersymmetric 
model can be extended by adding bilinear R-Parity violating interactions,
\begin{equation}
{\cal L}^{split}_{RpV} = \epsilon_i \tilde H_u^T i\sigma_2 L_i - \frac{1}{\sqrt{2}} a_i
H^T i\sigma_2 ( -\tilde g_d \sigma_j \tilde W^j + \tilde g'_d \tilde B ) L_i + h.c.
\end{equation}
The terms proportional to $\epsilon_i$ come from the usual bilinear interaction, 
while the terms proportional to $a_i$ appear after integrating out the sleptons
and retaining renormalizable operators~\cite{Diaz:2006ee}.
At this point, since higher order terms have been neglected, all the terms 
of the effective low energy Lagrangian are renormalizable (indicated by the subscript $r$).
The neutralinos and neutrinos mix themselves 
by means of the following mass matrix,
\begin{equation}
{\cal M}^{SS}_{r} = \left[\begin{array}{cc} {\mathrm M}_{\chi^0,r}^{SS} & (m^{SS}_r)^T \\ 
m^{SS}_r & 0 \end{array}\right],
\label{F0massmat}
\end{equation}
where the neutralino submatrix is given by
\begin{equation}
{\bf M}_{\chi^0,r}^{SS}=\left[\begin{array}{cccc}
M_1 & 0 & -\frac{1}{2}\tilde g'_d v & \frac{1}{2}\tilde g'_u v \\
0 & M_2 & \frac{1}{2}\tilde g_d v & -\frac{1}{2}\tilde g_u v \\
-\frac{1}{2}\tilde g'_d v & \frac{1}{2}\tilde g_d v & 0 & -\mu \\
\frac{1}{2}\tilde g'_u v & -\frac{1}{2}\tilde g_u v & -\mu & 0
\end{array}\right],
\label{X0massmat}
\end{equation}
and the mixing between neutralinos and neutrinos is defined by the submatrix,
\begin{equation}
m^{SS}_r=\left[\begin{array}{cccc}
-\frac{1}{2} \tilde g'_d a_1v & \frac{1}{2} \tilde g_d a_1v 
& 0 &\epsilon_1 \cr
-\frac{1}{2} \tilde g'_d a_2v & \frac{1}{2} \tilde g_d a_2v&0 
& \epsilon_2 \cr
-\frac{1}{2} \tilde g'_d a_3v & \frac{1}{2} \tilde g_d a_3v&0 
& \epsilon_3
\end{array}\right].
\end{equation}
Here, $v$ is the vacuum expectation value of the light Higgs field.
If one defines $\lambda_i=a_i \mu + \epsilon_i$ and block-diagonalizes the matrix in eq.~(\ref{F0massmat})
one obtains the well known effective neutrino mass matrix,
\begin{equation}
{\bf M}_\nu^{eff}=
\frac{v^2}{4\det{M_{\chi^0,r}^{SS}}}
\left(M_1 \tilde g_d^2 + M_2 \tilde g'^2_d \right)
\left[\begin{array}{cccc}
\lambda_1^2        & \lambda_1\lambda_2 & \lambda_1\lambda_3 \cr
\lambda_2\lambda_1 & \lambda_2^2        & \lambda_2\lambda_3 \cr
\lambda_3\lambda_1 & \lambda_3\lambda_2 & \lambda_3^2
\end{array}\right],
\label{treenumass}
\end{equation}
which gives the atmospheric neutrino mass scale. For notational purposes we call 
the matrix elements of this $3\times3$ 
neutrino mass matrix ${\bf M}_{\nu,ij}^{eff}=A\lambda_i\lambda_j$. 
The situation with Split Supersymmetry 
is that one loop corrections give $\Delta {\bf M}_{\nu,ij}^{eff}=(\Delta A) \lambda_i\lambda_j$, 
this is,
a term proportional to the tree level value. 
Thus, loop contributions in SS do not induce a second neutrino mass scale.
In the MSSM with BRpV something different happens,
since loop contributions additionally induce a term proportional to $\epsilon_i \epsilon_j$\cite{Hirsch:2000ef} 
and in accordance with the result in ref.~\cite{Hirsch:1997vz}. 
In order to obtain a second mass scale in Split Supersymmetry one alternative is to include 
extra contributions which go beyond the original SS or MSSM field content.
For example an extra contribution from gravity has been studied in \cite{Diaz:2009yz}. 
We consider here a different alternative, 
namely the effect of neglected contributions from non-renormalizable operators
that arise from the original MSSM Lagrangian.

Before working out the relevant non-renormalizable terms, we mention the fact that in Split Supersymmetry
the value of the light Higgs mass near $125$ GeV implies an upper bound on the scale $\widetilde m$
of the order of $10^5$-$10^6$ GeV \cite{Giudice:2004tc,Binger:2004nn}, in other words, the value of the light Higgs 
mass implies a rather low value of the Split SuSy scale.

%%%%%%%%%%%%%%%%%%%%%%%%%%%%%%%%%%%%%%%%%%%%%%%%%%%%%%%%%%%%%%%%%%%%%%%%%%%%%%%
\section{Split Supersymmetry beyond minimal integration of scalars}
%\subsection{non-renormalizable Operators}
%%%%%%%%%%%%%%%%%%%%%%%%%%%%%%%%%%%%%%%%%%%%%%%%%%%%%%%%%%%%%%%%%%%%%%%%%%%%%%%

Like in the previous section, the starting point is the MSSM Lagrangian
with the Higgs potential (\ref{HiggsPotMSSM}). However, now the integration 
of heavy scalars will not be truncated at the first contribution. For example, 
for the heavy Higgs the equation of motion is given by eq.~(\ref{HSS1}).
Instead of sending $\tilde{m}\rightarrow \infty$ which corresponds to simply 
erasing the $H_{SS}$ from the Lagrangian 
we now retain a  finite (but large)
value of $\tilde{m}$ and replace (\ref{HSS1}) back into the Lagrangian~(\ref{HiggsPotMSSM}). 
In this way, new operators arise as 
as a consequence of the finite mass of the heavy scalars. 
One of those terms is
\begin{equation}
{\cal L}^{nr} 
%\supset 
\owns \frac{1}{8} (g^2+g'^2) \frac{c^2_\alpha}{\widetilde m^4} \left\{
                  H_{SM}^T i\sigma_2 \left[ g\left( \sigma \cdot \widetilde{W} \right)\, -\, g' \widetilde{B} \right]\widetilde{H}_d
                 \right\}^2,
\label{nonRL}
\end{equation}
%
%%\frac{1}{8} (g^2+g'^2) \frac{c^2_\alpha}{\widetilde m^4} 
%%H^\dagger_{SM} H_{SM} \left\{ 
%%g^2 \left[ (\sigma\cdot\widetilde W)\widetilde H_d \right]^2 +
%%g'^{\,2}\left(\widetilde B\widetilde H_d  \right)^2 \right\} + ...
%[MAD, MINOR CHANGES IN THE FORMULA] 
which is a non-renormalizable dimension 8 operator (from now on, the script $nr$ refers to
``non-renormalizable''). The value of the coupling 
is given at $\tilde m$, and it leads to a neutrino mass contribution that will 
be developed in the next section. Notice that the bilinear R-parity violating
mixing in matrix (\ref{F0massmat}) allows us to convert from a $\widetilde{H}_d$
to a neutrino after block diagonalization \cite{Hirsch:2000ef,Diaz:2003as}.
Analogously one finds the following $d=7$ operator when integrating out 
down squarks
\begin{eqnarray}
%\mathcal{L}^{nr}_{sd} &=& 
\mathcal{L}^{nr} &\owns&
-\frac{A_d Y_d^3}{m_{\widetilde{Q}}^2 m_{\widetilde{d}}^2} \left( c_\alpha H_{SM}^T i\sigma_2 \overline{\widetilde{H}}_d d_R\right) \left( \overline{Q} i\sigma_2 \widetilde{H}^c_d\right)\,+\, h.c. \label{eq:dsdnr}
\end{eqnarray}
%(DOUBLE CHECK).
Similar operators appear after integrating
out sleptons 
and the procedure invoked is the same (although with small differences). 
These two kinds of operators (\ref{nonRL},\ref{eq:dsdnr}) will induce corrections
that have been claimed to be, jointly with the neutralino-neutrino corrections,
the largest contributions to the solar mass difference \cite{Diaz:2003as,Diaz:2014jta}. 
Notice that in order to write down a similar expression for the sleptons we just need
to do the replacement : $d_R \rightarrow e_R$, $Q_i \rightarrow L_i$, and the 
subindex $d \rightarrow e$ (but not for $\widetilde{H}_d$).

In order to write down this last coupling, we perform the decoupling of squarks
in the soft trilinear sector of the MSSM instead the Higgs potential, as we
did with the $d=8$ operator (\ref{nonRL}). In the MSSM this term allows
us to write the down-sdown loop. Nevertheless, equation (\ref{eq:dsdnr}) 
(and the sleptonic analogue) introduce a dependence on the trilinear scalar 
couplings $A_d$ ($A_e$, for sleptons). 
%Later we explore the loops induced by these operators.

The non-renormalizable operators given in eqs.~(\ref{nonRL}) and (\ref{eq:dsdnr}) lay the seed for
the generation of the needed neutrino masses due to quantum corrections.
This mechanism will be explained in the following section.

%%%%%%%%%%%%%%%%%%%%%%%%%%%%%%%%%%%%%%%%%%%%%%%%%%%%%%%%%%%%%%%%%%%%
\section{Loop Corrections}
%%%%%%%%%%%%%%%%%%%%%%%%%%%%%%%%%%%%%%%%%%%%%%%%%%%%%%%%%%%%%%%%%%%%

\subsection{Calculating the Loops}
The non-renormalizable terms in the Lagrangian given in eq.~(\ref{nonRL}) lead to the following graph 
contributing to the neutrino mass matrix,
\begin{center}
\vspace{-50pt} \hfill \\
\begin{picture}(130,100)(40,45) % y_2 controls equation position
\SetWidth{1.0}
\Line(0,50)(40,50)
\Line(40,50)(80,50)
\BCirc(40,50){3}
\Line(120,50)(80,50)
\Line(160,50)(120,50)
\BCirc(120,50){3}
\CArc(80,25)(25,0,360)
\DashLine(60,80)(80,50){4}
\DashLine(100,80)(80,50){4}
\Text(56.3,77.2)[lb]{$\times$}
\Text(96.3,77.2)[lb]{$\times$}
\Text(5,55)[lb]{$\nu_i$}
\LongArrow(30,70)(38,53.2)
\Text(25,70)[lb]{$\frac{\epsilon_i}{\mu}$}
\Text(50,55)[lb]{$\widetilde H_d$}
\Text(53,86)[lb]{$\frac{v}{\sqrt{2}}$}
\Text(95,86)[lb]{$\frac{v}{\sqrt{2}}$}
\Text(95,55)[lb]{$\widetilde H_d$}
\LongArrow(130,70)(122,53.2)
\Text(125,70)[lb]{$\frac{\epsilon_j}{\mu}$}
\Text(145,55)[lb]{$\nu_j$}
\Text(68,-16)[lb]{$\widetilde W,\widetilde B$}
\CCirc(80,50){3}{1}{1}
\end{picture}
$
$
\vspace{30pt} \hfill \\
\end{center}
\vspace{40pt}
The non-renormalizable vertex is represented in the graph by a full circle at the center.
The open circles represent the mixing between the neutrinos and the down type higgsino
due to R-Parity violation. 
The central vertex of this diagram arises from the corresponding diagram in the underlying supersymmetric theory
shown in the figure below\\
\begin{center}
\vspace{-50pt} \hfill \\
\begin{picture}(330,100)(40,45) % y_2 controls equation position
\SetWidth{1.0}
\Line(0,50)(40,50)
\Line(40,50)(10,20)
\DashLine(40,50)(70,80){4}
\DashLine(70,80)(100,50){4}
\Line(100,50)(140,50)
\Line(100,50)(130,20)
\DashLine(40,110)(70,80){4}
\DashLine(70,80)(100,110){4}
\Text(15,110)[lb]{$H_{SM}$}
\Text(105,110)[lb]{$H_{SM}$}
\Text(32,67)[lb]{$H_{SS}$}
\Text(88,67)[lb]{$H_{SS}$}
\Text(0,55)[lb]{$\widetilde H_d$}
\Text(130,55)[lb]{$\widetilde H_d$}
\Text(0,5)[lb]{$\widetilde W,\widetilde B$}
\Text(120,5)[lb]{$\widetilde W,\widetilde B$}
\LongArrow(180,50)(200,50)
\Line(240,50)(280,50)
\Line(320,50)(280,50)
\DashLine(260,80)(280,50){4}
\DashLine(300,80)(280,50){4}
\Text(230,55)[lb]{$\widetilde H_d$}
\Text(245,86)[lb]{$H_{SM}$}
\Text(298,86)[lb]{$H_{SM}$}
\Text(320,55)[lb]{$\widetilde H_d$}
\CCirc(280,50){3}{1}{1}
\Line(310,20)(280,50)
\Line(250,20)(280,50)
\Text(235,5)[lb]{$\widetilde W,\widetilde B$}
\Text(295,5)[lb]{$\widetilde W,\widetilde B$.}
\end{picture}
$
$
\vspace{30pt} \hfill \\
\end{center}
The SM Higgs field acquires a vacuum expectation value $v$. Inside the loop we can have winos 
or binos. The mass contributions of those two loops can be summarized as,
\begin{equation}
%\Delta M_{\nu (\chi)}^{nr} 
\Delta M_{\nu}^{nr} 
= \frac{1}{256\pi^2} (g^2+g'^2) \frac{v^2 s_\beta^2}{\widetilde m^4}
\left[ g^2 M_2 \tilde{A}_0(M_2^2) + g'^2 M_1 \tilde{A}_0(M_1^2) \right] \frac{\epsilon_i\epsilon_j}{\mu^2},
\label{NonRenCont}
\end{equation}
where we see the dimension $8$ vertex, the two vevs of the light Higgs field, the mixing between 
neutrinos and the down type higgsino, the loop factor and the finite Veltman's function $\tilde A_0$. 
Since the solar neutrino 
mass scale is very small, the contribution from (\ref{NonRenCont}) can be important,
even though it is suppressed by a factor of $1/{\tilde m}^4$. 
An estimation of the orders of magnitudes in
eq.~(\ref{NonRenCont}) confirms that this contribution can do the job. 
In the next subsection this contribution will be evaluated numerically.

As it was indicated in the previous section, there are two other contributions 
that we studied in detail: A loop with down quarks, and a loop of leptons which 
are produced after the decoupling of down squarks and sleptons. In these 
two cases, the contribution can be related in a similar way to the diagram: 
\begin{center}
\vspace{-50pt} \hfill \\
\begin{picture}(130,100)(40,45) % y_2 controls equation position
\SetWidth{1.0}
\Line(0,50)(40,50)
\Line(40,50)(80,50)
\BCirc(40,50){3}
\Line(120,50)(80,50) 
\Line(160,50)(120,50)
\BCirc(120,50){3}
\CArc(80,25)(25,0,360)
%$$%\DashLine(60,80)(80,50){4}
%$$%\DashLine(100,80)(80,50){4}
%$$%\Text(56.3,77.2)[lb]{$\times$}
%$$%\Text(96.3,77.2)[lb]{$\times$}
\Text(5,55)[lb]{$\nu_i$}
\LongArrow(30,70)(38,53.2)
\Text(25,70)[lb]{$\frac{\epsilon_i}{\mu}$}
\Text(50,55)[lb]{$\widetilde H_d$}
\Text(80,86)[lb]{$\frac{v}{\sqrt{2}}$}
\Text(95,55)[lb]{$\widetilde H_d$}
\LongArrow(130,70)(122,53.2)
\Text(125,70)[lb]{$\frac{\epsilon_j}{\mu}$}
\Text(145,55)[lb]{$\nu_j$}
%$$%\Text(68,-16)[lb]{$\widetilde W,\widetilde B$}
\Text(35,20)[lb]{$d_R$}
\Text(110,20)[lb]{$\overline{Q}$}
\CCirc(80,50){3}{1}{1}
%%\CCirc(80,-0.5){3}{1}{1}
\DashLine(80,50)(80,80){4}
\Text(76.0,78)[lb]{$\times$}
\DashLine(80,-0.5)(80,-30.5){4}
\Text(76.0,-33.5)[lb]{$\times$}
\Text(82,-30.5)[lb]{$\frac{v}{\sqrt{2}}$}.
\end{picture}
$
$
\vspace{30pt} \hfill \\
\end{center}
\vspace{40pt}
The corresponding  $d=7$ operator [see eq.~(\ref{eq:dsdnr})] gives rise to a numerically irrelevant 
contribution to the neutrino mass matrix. This happens despite of the fact that the bottom-sbottom loops are in many cases 
important in the usual MSSM+BRpV models. The reason is that when sbottom quarks are decoupled with mass $\tilde{m}$, the contribution 
from the corresponding non-renormalizable operator becomes irrelevant due to the smallness of the bottom 
quark mass compared to the mass of the gauginos.\\

In summary, in order to write down the corrections for neutrino masses, we use 
the notation for the $3\times3$ neutrino corrected mass matrix in the form of
\begin{equation}
{\bf M}_\nu^{eff} = A\lambda_i\lambda_j + C\epsilon_i\epsilon_j \label{eq:genformneut}.
\end{equation}
with,
\begin{eqnarray}
A&=&\frac{v^2}{4\det{M_{\chi^0,r}^{SS}}}\left(M_1 \tilde g_d^2 + M_2 \tilde g'^2_d \right)
\nonumber\\
C&=&\frac{1}{256\pi^2} (g^2+g'^2) \frac{v^2 s_\beta^2}{\widetilde m^4 \mu^2}
\left[ g^2 M_2 \tilde{A}_0(M_2^2) + g'^2 M_1 \tilde{A}_0(M_1^2) \right]
\end{eqnarray}
as can be read from eqs.~(\ref{treenumass}) and (\ref{NonRenCont}).
It is by virtue of a the second term that one can expect an additional non-vanishing neutrino 
mass scale. 

Notice that the contribution in eq.~(\ref{NonRenCont}) 
is in principle not finite (if we imagine replacing $\tilde A_0$ by $A_0$), 
as opposite to the renormalizable case. However,
there is no reason for concern here, since the inclusion of a counterterm 
that absorbs this infinity must be considered after the decoupling of the
heavy degrees of freedom. A similar case, where loops finite in a 
renormalizable theory, are turned into infinite in the effective theory 
can be found at the reference \cite{Buchalla:1995vs}. In addition, it is 
worthwhile to mention that value of the scale $\tilde{m}$ between $10^{3}\, -\, 10^{5}$ GeV 
helps to realize the radiative mechanism for neutrino masses when one wants to compute the 
Wilson coefficients for Weinberg operator via new degrees of freedom 
added to the SM~\cite{Sierra:2016rcz}.

%%%%%%%%%%%%%%%%%%%%%%%%%%%%%%%
\subsection{Numerical Results}
%%%%%%%%%%%%%%%%%%%%%%%%%%%%%%%

Our intention is to explore whether the non-renormalizable operators account 
for neutrino physics or their scale suppression makes the neutrino phenomenology 
unfeasible. Therefore, if a positive result is found, it is sufficient to perform a scan in a window of the
parameter space.
In order to see how a solar squared mass difference arises from the effect 
of the loop, we computed the neutrino squared mass differences and the 
corresponding mixing angles by using the equation (\ref{eq:genformneut}). The
atmospheric square mass difference can be generated via the $\lambda$ 
parameters as it happens in the models that consider bilinearly violated 
R-parity.

We also implemented the condition that a Higgs mass is within the zone allowed by the 
experiments~\cite{Aad:2015zhl}. This point is crucial since, as it was indicated above, the 
Higgs mass is one of the most stringent parameters in order to define the 
scale where new physics appears. It  turns out that this conditions 
restricts the SuSy mass scale to be of the order of $\tilde m \sim 10^{3}- 10^{5}$ GeV
in order to avoid an extreme fine tuning. We also imposed the condition that the spectrum 
of supersymmetric particles fulfills the experimental constraints 
from the supersymmetric searches~\cite{Aaboud:2016zpr}. The ranges where we varied the parameters 
of the model are depicted on tables \ref{tab:scanI} and \ref{tab:scanII}.

\begin{table}
\begin{center}
\begin{tabular}{c@{\hskip 1.0cm}c@{\hskip 1.0cm}c}
Value & Min. $[\,GeV\,]$ & Max. $[\,GeV\,]$ \\
\hline
\hline
$M_\chi$ &      500      & 1500 \\
$M_1$    & $M_\chi$ & $1.5\,M_\chi$ \\
$M_2$    & $M_\chi$ & $1.5\,M_\chi$ \\
$M_3$    & $10^{3.5}$ & $10^{5}$ \\
$\mu$    & $M_\chi$ & $1.5\,M_\chi$ \\
$\epsilon_1$ & $-10.0$ & $10.0$ \\
$\epsilon_2$ & $-10.0$ & $10.0$ \\
$\epsilon_3$ & $-10.0$ & $10.0$ \\
$\tilde{m}$  & $10^{3.5}$ & $10^{5}$
\end{tabular}
\caption{Ranges for parameters with mass units.}
\label{tab:scanI}\vspace{1.0cm}
\begin{tabular}{c@{\hskip 1.0cm}c@{\hskip 1.0cm}c}
Value & Min. & Max. \\
\hline
\hline
$\tan \beta$ & $1$ & $45$ \\
$\lambda_1$ & $-1.0$ & $1.0$ \\
$\lambda_2$ & $-1.0$ & $1.0$ \\
$\lambda_3$ & $-1.0$ & $1.0$
\end{tabular}
\caption{Ranges for parameters without units.}\label{tab:scanII}
\end{center}
\end{table}

From the table (\ref{tab:scanI}), we can highlight that the value of $M_3$,
this is, the Gluino mass, can be safely put at high scales, since the
Gluino does not interact with SM particles in this context (the squarks
have been decoupled), there is no reason to keep it at low scales. 
Two comments about the $\lambda_i$
parameters are at place here. 
First, on the contrary to the notation of other BRpV models, in this
case the $\lambda_i$ have no units. 
Second, the values of $\lambda_i$ have been chosen in order to fulfill
the atmospheric square mass difference. 
This was achieved by choosing the parameters once the $A$ factor on eq. (\ref{eq:genformneut})
has been computed.
When performing the scan, we obtained the full spectrum of particles which
is in agreement with the constraints from the searches for supersymmetric particles\cite{Aad:2012tfa}.
The limiting values of the obtained spectra are shown in the table 
\ref{tab:spectrum}. As mentioned above, it is imposed to have a spectrum 
where the masses of neutralinos and charginos are below the scale $\tilde{m}$,
otherwise it would be inconsistent to integrate out other particles with masses
of the order of $\tilde m$.
\begin{table}
\begin{center}
\begin{tabular}{c@{\hskip 1.0cm}c@{\hskip 1.0cm}c@{\hskip 1.0cm}c@{\hskip 1.0cm}c}
Parameter & & Min. Value & Max. Value & Units\\
\hline
\hline
lightest neutralino          & & $5.10\cdot 10^{2}$                 & $1.98\cdot 10^{3}$                  &  $GeV$   \\
higgs mass                   & & $1.24\cdot 10^{2}$                 & $1.26\cdot 10^{2}$                  &  $GeV$   \\
Gluino mass                  & & $3.46\cdot 10^{3}$                 & $6.54\cdot 10^{4}$                  &  $GeV$   \\
tan$\beta$                   & & $2.10\cdot 10^{0}$                 & $4.88\cdot 10^{1}$                  &  $-$     \\
neutrino Physics             & $\Delta m_{\odot}^2$      & $6.04\cdot 10^{-5}$ & $8.93\cdot 10^{-5}$      &  $eV^2$  \\
                             & $\Delta m_{atm}^2$        & $2.44\cdot 10^{-3}$ & $2.48\cdot 10^{-3}$      &  $eV^2$  \\
                             & $\sin^2 \theta_{\odot}$   & $2.81\cdot 10^{-1}$ & $3.72\cdot 10^{-1}$      &  $-$     \\
                             & $\sin^2 \theta_{atm}$     & $5.39\cdot 10^{-1}$ & $5.86\cdot 10^{-1}$      &  $-$     \\
                             & $\sin^2 \theta_{rea}$     & $1.80\cdot 10^{-2}$ & $2.77\cdot 10^{-2}$      &  $-$
\end{tabular}
\caption{Some values for the parameters obtained in the search.}\label{tab:spectrum}
\end{center}
\end{table}

\begin{figure}[ht]
\centerline{\protect\vbox{\epsfig{file=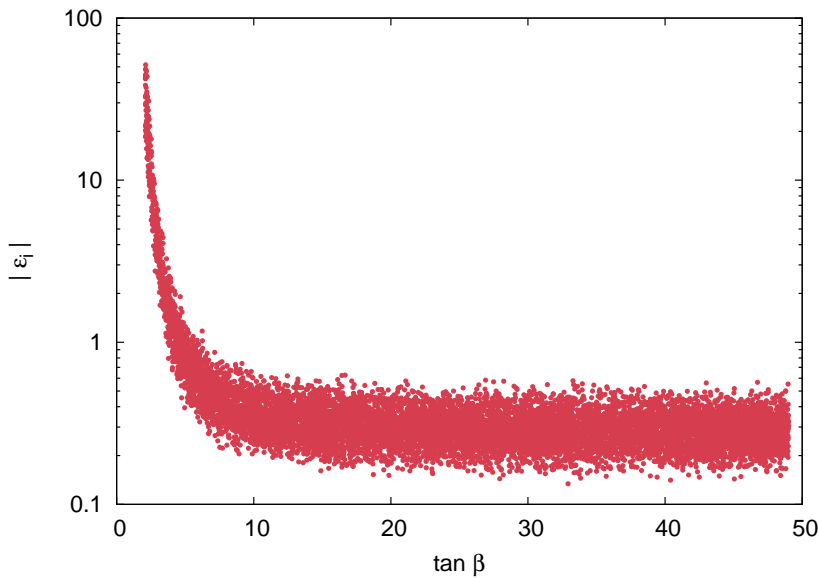,width=0.4\textwidth,angle=-90}\vspace{0.5cm}
}}
\caption{\it Modulus of $\vec{\epsilon}$ as a function of $\tan\beta$. The 
strong correlation between the parameters, as the accumulation of points
for  $0.1\, \leq\, \left|\vec{\epsilon}\right|\, \leq\, 0.5$  at
$\tan \beta\,\geq\, \sim 10$ is given by neutrino physics (see body of the text).}
\label{tanbvse}
\end{figure}

The main result of this section is that the parameter space allows to meet
the neutrino and Higgs requirement and the collider bounds on SuSy masses. 
One notices a strong correlation 
between the parameters, but the good scenarios are not accumulated around 
an isolated point in the parameter space. 
Fig.~\ref{tanbvse} shows $|\vec\epsilon\,|$ as a function of $\tan\beta$. 
One notes that the points indicating allowed parameter space show a 
balance between $|\vec\epsilon\,|$ and $\tan\beta$. 
This can be understood 
from the fact that the presence of $s_{\beta}^2$ in eq.~(\ref{NonRenCont}) 
implies that larger values of $|\vec\epsilon\,|$ are needed at low 
$\tan\beta$ to obtain the correct solar mass scale. Notice that in this 
analysis, the $\epsilon$ values lie upon a range which is taken to be 
natural, since it should remain small in order to give rise to neutrino 
masses. Notice further that in the sector of large $\tan \beta$ the value of 
$\epsilon$ saturates around $10^{-1}$.
This can be understood from the observation that
 $\sin \beta <1$.

%
%%%%%%%%%% FIGURE %%%%%%%%%%%%%%%%%%
\begin{figure}[ht]
\centerline{\protect\vbox{\epsfig{file=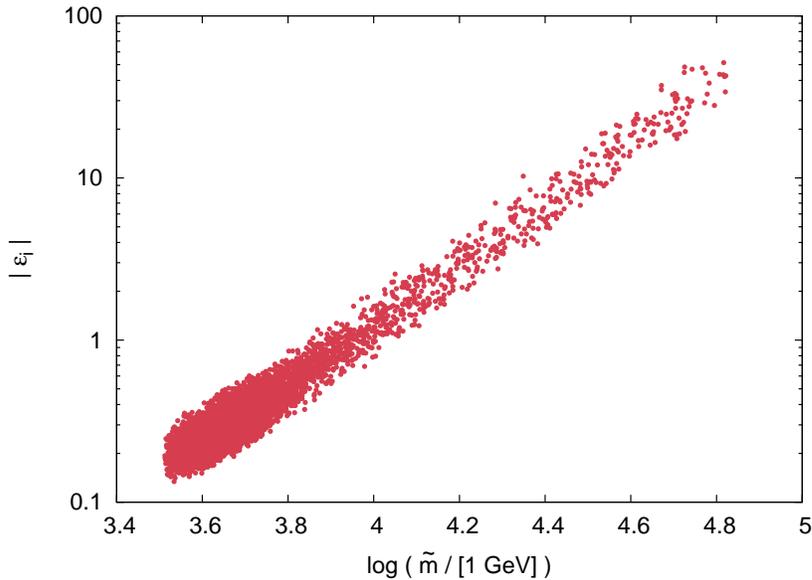,width=0.4\textwidth,angle=-90}\vspace{0.5cm}
}}
\caption{\it Modulus of $\vec{\epsilon}$ as a function of $\widetilde{m}$.
The accumulation of points between $3.5\,\leq\,\widetilde{m}\,\leq\,4.0$ obeys to neutrino physics, which
pushes to have $\left| \vec{\epsilon} \right|\, \sim\, 0.25$ 
%(see body of the text and fig.~\ref{tanbvse}) 
}
\label{evslmtilde}
\end{figure}

%%%%%%%%%%%%%%%%%%%%%%%%%%%%%%%%%%%%
%
%
The fig.~\ref{evslmtilde} shows $|\vec\epsilon\,|$ as a function of the SS scale $\widetilde m$.
There, one sees that the allowed parameter space follows a nice linear relation 
in the logarithmic plot, where a growth of one order of magnitude in 
$\tilde m$ is compensated by a growth of two orders of magnitude in 
$|\vec\epsilon\,|$. This can be understood from relation (\ref{NonRenCont}), 
since $\Delta M_{\nu}^{nr}\sim \epsilon_i\epsilon_j/\tilde m^4$. The 
fact that this correlation between the $|\vec\epsilon\,|$ and  $\tilde m$ 
does not extend to arbitrary values of  $\tilde m$ is due the bounds imposed 
from the Higgs sector.

%
%%%%%%%%%% FIGURE %%%%%%%%%%%%%%%%%%
%%%%%%%%%%%%%%%%%%%%%%%%%%%%%%%%%%%%
%

%%%%%%%%%%%%%%%%%%%%%%%%%%%%%%%%%%%%%%%%%%%%%%%%%%%%%%%%%%%%%%%%%%%%%%%%%%%%%%%
\section{Summary}
%%%%%%%%%%%%%%%%%%%%%%%%%%%%%%%%%%%%%%%%%%%%%%%%%%%%%%%%%%%%%%%%%%%%%%%%%%%%%%%
%
This paper is dedicated to the 
tension between
Split SuSy on the one hand, 
a good candidate for a low energy MSSM in terms of collider physics, 
and the same model
on the other hand, not doing well 
for neutrino physics since it cannot account for the solar mass 
difference of neutrinos. We explore the possibility that the problem of Split SuSy with the solar
mass difference originates from the fact that in the transition from the MSSM
to Split SuSy the scalar fields are integrated out by a leading order approximation only.

In order to study this working hypothesis we include further terms
in the integrating out procedure, which have been previously neglected.
Those terms appear in the effective low energy Split SuSy Lagrangian
in the form of non-renormalizable operators.
With those inclusions we calculate quantum corrections and it is found that 
indeed a non-trivial contribution to the neutrino mass matrix is generated 
after spontaneous symmetry breaking.
It is found that in particular the contribution coming from (\ref{NonRenCont})
has the potential of generating the observed solar mass difference.
Finally, it is shown that this extended version of Split SuSy
is indeed capable of reproducing the observed neutrino oscillations
by simultaneously avoiding a strong fine tuning of the Higgs mass,
if the mass scale $\tilde{m}$ is rather moderate $\sim 10^4$~GeV.

The finding of good neutrino phenomenology at such a moderate scale $\tilde{m}$ further
supports our working hypothesis, namely, that terms that have been
previously neglected in the integration out procedure can actually re-conciliate
Split SuSy with the solar neutrino mass difference.

%%%%%%%%%%%%%%%%%%%%%%%%%%%%%%%%%%%%%%%%%%%%%%%%%%%%%%%%%%%%%%%%%
%%%%%%%%%%%%%%%%%%%%%%%%%%%%%%%%%%%%%%%%%%%%%%%%%%%%%%%%%%%%%%%%%%%%

%%%%%%%%%%%%%%%%%%%%%%%%%%%%%%%%%%%%%%%%%%%%%%%%%%%%%%%%%%%%%%%%%%
\begin{acknowledgments}
{\small 
This work was partly funded by the Spanish grants FPA2014-58183-P, Multidark CSD2009-00064, 
SEV-2014-0398 (MINECO) and PROMETEOII/2014/084 (Generalitat Valenciana). NR was funded 
by becas de postdoctorado en el extranjero Conicyt/Becas Chile 74150028 and also wants 
to thank B. Panes, M. Platscher and A. Vicente for useful discussions and comments,
and the people at PUC-Ch, in particular, the coauthors of this article for the hospitality.
BK acknowledges the support of Fondecyt 1161150 and MAD Fondecyt 1141190.
}
\end{acknowledgments}
%%%%%%%%%%%%%%%%%%%%%%%%%%%%%%%%%%%%%%%%%%%%%%%%%%%%%%%%%%%%%%%%%

%%%%%%%%%%%%%%%%%%%%%%%%%%%%%%%%%%%%%%%%%%%%%%%%%%%%%%%%%%%%%%%%%%%%%%%%%%%%%%


\begin{thebibliography}{99}
%%%%%%%%%%%%%%%%%%%%%%%%%%%%%%%%%%%%%%%%%%%%%%%%%%%%%%%%%%%%%%%%%%%%%%%%%%%%%%

%\cite{Aad:2012tfa}
\bibitem{Aad:2012tfa} 
  G.~Aad {\it et al.} [ATLAS Collaboration],
  %``Observation of a new particle in the search for the Standard Model Higgs boson with the ATLAS detector at the LHC,''
  Phys.\ Lett.\ B {\bf 716}, 1 (2012)
  doi:10.1016/j.physletb.2012.08.020
  [arXiv:1207.7214 [hep-ex]].
  %%CITATION = doi:10.1016/j.physletb.2012.08.020;%%
  %6078 citations counted in INSPIRE as of 30 May 2016
  
%\cite{Chatrchyan:2012xdj}
\bibitem{Chatrchyan:2012xdj} 
  S.~Chatrchyan {\it et al.} [CMS Collaboration],
  %``Observation of a new boson at a mass of 125 GeV with the CMS experiment at the LHC,''
  Phys.\ Lett.\ B {\bf 716}, 30 (2012)
  doi:10.1016/j.physletb.2012.08.021
  [arXiv:1207.7235 [hep-ex]].
  %%CITATION = doi:10.1016/j.physletb.2012.08.021;%%
  %5934 citations counted in INSPIRE as of 30 May 2016
  


%$$%%\cite{Davis:1968cp}
%$$%\bibitem{Davis:1968cp}
%$$%  R.~Davis, Jr., D.~S.~Harmer and K.~C.~Hoffman,
%$$%  %``Search for neutrinos from the sun,''
%$$%  Phys.\ Rev.\ Lett.\  {\bf 20} (1968) 1205.
%$$%  doi:10.1103/PhysRevLett.20.1205
%$$%  %%CITATION = doi:10.1103/PhysRevLett.20.1205;%%
%$$%  %1175 citations counted in INSPIRE as of 16 Jun 2016
%$$%  
%$$%%\cite{Fukuda:1998mi}
%$$%\bibitem{Fukuda:1998mi}
%$$%  Y.~Fukuda {\it et al.} [Super-Kamiokande Collaboration],
%$$%  %``Evidence for oscillation of atmospheric neutrinos,''
%$$%  Phys.\ Rev.\ Lett.\  {\bf 81} (1998) 1562
%$$%  doi:10.1103/PhysRevLett.81.1562
%$$%  [hep-ex/9807003].
%$$%  %%CITATION = doi:10.1103/PhysRevLett.81.1562;%%
%$$%  %4841 citations counted in INSPIRE as of 16 Jun 2016

%\cite{Weinberg:1979sa}
\bibitem{Weinberg:1979sa}
  S.~Weinberg,
  %``Baryon and Lepton Nonconserving Processes,''
  Phys.\ Rev.\ Lett.\  {\bf 43} (1979) 1566.
  doi:10.1103/PhysRevLett.43.1566
  %%CITATION = doi:10.1103/PhysRevLett.43.1566;%%
  %1169 citations counted in INSPIRE as of 18 Jun 2016
  
%%%\cite{An:2012eh}
%%\bibitem{An:2012eh}
%%  F.~P.~An {\it et al.} [Daya Bay Collaboration],
%%  %``Observation of electron-antineutrino disappearance at Daya Bay,''
%%  Phys.\ Rev.\ Lett.\  {\bf 108} (2012) 171803
%%  doi:10.1103/PhysRevLett.108.171803
%%  [arXiv:1203.1669 [hep-ex]].
%%  %%CITATION = doi:10.1103/PhysRevLett.108.171803;%%
%%  %1543 citations counted in INSPIRE as of 16 Jun 2016  

%\cite{Zee:1985id}
\bibitem{Zee:1985id}
  A.~Zee,
  %``Quantum Numbers of Majorana Neutrino Masses,''
  Nucl.\ Phys.\ B {\bf 264} (1986) 99.
  doi:10.1016/0550-3213(86)90475-X
  %%CITATION = doi:10.1016/0550-3213(86)90475-X;%%
  %298 citations counted in INSPIRE as of 24 Jun 2016  
%\cite{Babu:1988ki}
%\bibitem{Babu:1988ki}
  K.~S.~Babu,
  %``Model of 'Calculable' Majorana Neutrino Masses,''
  Phys.\ Lett.\ B {\bf 203} (1988) 132.
  doi:10.1016/0370-2693(88)91584-5
  %%CITATION = doi:10.1016/0370-2693(88)91584-5;%%
  %484 citations counted in INSPIRE as of 24 Jun 2016
  
%\cite{Aoki:2008av}
\bibitem{Aoki:2008av}
  M.~Aoki, S.~Kanemura and O.~Seto,
  %``Neutrino mass, Dark Matter and Baryon Asymmetry via TeV-Scale Physics without Fine-Tuning,''
  Phys.\ Rev.\ Lett.\  {\bf 102} (2009) 051805
  doi:10.1103/PhysRevLett.102.051805
  [arXiv:0807.0361 [hep-ph]].
  %%CITATION = doi:10.1103/PhysRevLett.102.051805;%%
  %212 citations counted in INSPIRE as of 24 Jun 2016
%\cite{Gustafsson:2012vj}
%\bibitem{Gustafsson:2012vj}
  M.~Gustafsson, J.~M.~No and M.~A.~Rivera,
  %``Predictive Model for Radiatively Induced Neutrino Masses and Mixings with Dark Matter,''
  Phys.\ Rev.\ Lett.\  {\bf 110} (2013) no.21,  211802
   Erratum: [Phys.\ Rev.\ Lett.\  {\bf 112} (2014) no.25,  259902]
  doi:10.1103/PhysRevLett.110.211802, 10.1103/PhysRevLett.112.259902
  [arXiv:1212.4806 [hep-ph]].
  %%CITATION = doi:10.1103/PhysRevLett.110.211802, 10.1103/PhysRevLett.112.259902;%%
  %91 citations counted in INSPIRE as of 18 Jun 2016

%\cite{Aad:2015baa}
\bibitem{Aad:2015baa} 
  G.~Aad {\it et al.} [ATLAS Collaboration],
  %``Summary of the ATLAS experimentâs sensitivity to supersymmetry after LHC Run 1 â interpreted in the phenomenological MSSM,''
  JHEP {\bf 1510}, 134 (2015)
  doi:10.1007/JHEP10(2015)134
  [arXiv:1508.06608 [hep-ex]];
  %%CITATION = doi:10.1007/JHEP10(2015)134;%%
  %13 citations counted in INSPIRE as of 12 Jan 2016
%\cite{Khachatryan:2015vra}
%\bibitem{Khachatryan:2015vra} 
  V.~Khachatryan {\it et al.} [CMS Collaboration],
  %``Searches for Supersymmetry using the M$_{T2}$ Variable in Hadronic Events Produced in pp Collisions at 8 TeV,''
  JHEP {\bf 1505}, 078 (2015)
  doi:10.1007/JHEP05(2015)078
  [arXiv:1502.04358 [hep-ex]].
  %%CITATION = doi:10.1007/JHEP05(2015)078;%%
  %40 citations counted in INSPIRE as of 12 Jan 2016
  
%\cite{ArkaniHamed:2004fb}
\bibitem{ArkaniHamed:2004fb}
  N.~Arkani-Hamed and S.~Dimopoulos,
  %``Supersymmetric unification without low energy supersymmetry and signatures for fine-tuning at the LHC,''
  JHEP {\bf 0506} (2005) 073
  [hep-th/0405159].
  %%CITATION = HEP-TH/0405159;%%
  %794 citations counted in INSPIRE as of 21 Jan 2015

%\cite{Giudice:2004tc}
\bibitem{Giudice:2004tc}
  G.~F.~Giudice and A.~Romanino,
  %``Split supersymmetry,''
  Nucl.\ Phys.\ B {\bf 699} (2004) 65
   [Erratum-ibid.\ B {\bf 706} (2005) 65]
  [hep-ph/0406088].
  %%CITATION = HEP-PH/0406088;%%
  %591 citations counted in INSPIRE as of 21 Jan 2015

%\cite{Binger:2004nn}
\bibitem{Binger:2004nn}
  M.~Binger,
  %``Higgs boson mass in split supersymmetry at two-loops,''
  Phys.\ Rev.\ D {\bf 73} (2006) 095001
  doi:10.1103/PhysRevD.73.095001
  [hep-ph/0408240].
  %%CITATION = doi:10.1103/PhysRevD.73.095001;%%
  %53 citations counted in INSPIRE as of 19 Jun 2016
  
%\cite{Diaz:2006ee}
\bibitem{Diaz:2006ee} 
  M.~A.~Diaz, P.~Fileviez Perez and C.~Mora,
  %``Neutrino Masses in Split Supersymmetry,''
  Phys.\ Rev.\ D {\bf 79}, 013005 (2009)
  doi:10.1103/PhysRevD.79.013005
  [hep-ph/0605285].
  %%CITATION = doi:10.1103/PhysRevD.79.013005;%%
  %13 citations counted in INSPIRE as of 12 Jan 2016
  
%\cite{Barbieri:1979hc}
\bibitem{Barbieri:1979hc} 
  R.~Barbieri, J.~R.~Ellis and M.~K.~Gaillard,
  %``Neutrino Masses and Oscillations in SU(5),''
  Phys.\ Lett.\ B {\bf 90}, 249 (1980).
  doi:10.1016/0370-2693(80)90734-0
  %%CITATION = doi:10.1016/0370-2693(80)90734-0;%%
  %194 citations counted in INSPIRE as of 07 Jul 2016
  %\cite{Berezinsky:2004zb}
%\cite{Akhmedov:1992hh}
%\bibitem{Akhmedov:1992hh} 
  E.~K.~Akhmedov, Z.~G.~Berezhiani and G.~Senjanovic,
  %``Planck scale physics and neutrino masses,''
  Phys.\ Rev.\ Lett.\  {\bf 69}, 3013 (1992)
  doi:10.1103/PhysRevLett.69.3013
  [hep-ph/9205230].
  %%CITATION = doi:10.1103/PhysRevLett.69.3013;%%
  %185 citations counted in INSPIRE as of 07 Jul 2016
%\cite{deGouvea:2000jp}
%\bibitem{deGouvea:2000jp} 
  A.~de Gouvea and J.~W.~F.~Valle,
  %``Minimalistic neutrino mass model,''
  Phys.\ Lett.\ B {\bf 501}, 115 (2001)
  doi:10.1016/S0370-2693(01)00103-4
  [hep-ph/0010299].
  %%CITATION = doi:10.1016/S0370-2693(01)00103-4;%%
  %42 citations counted in INSPIRE as of 07 Jul 2016
%\bibitem{Berezinsky:2004zb} 
  V.~Berezinsky, M.~Narayan and F.~Vissani,
  %``Low scale gravity as the source of neutrino masses?,''
  JHEP {\bf 0504}, 009 (2005)
  doi:10.1088/1126-6708/2005/04/009
  [hep-ph/0401029].
  %%CITATION = doi:10.1088/1126-6708/2005/04/009;%%
  %10 citations counted in INSPIRE as of 07 Jul 2016
  
%\cite{Diaz:2009yz}
\bibitem{Diaz:2009yz}
  M.~A.~Diaz, B.~Koch and B.~Panes,
  %``Gravity Effects on Neutrino Masses in Split Supersymmetry,''
  Phys.\ Rev.\ D {\bf 79} 113009 (2009) 
  doi:10.1103/PhysRevD.79.113009
  [arXiv:0902.1720 [hep-ph]].
  %%CITATION = doi:10.1103/PhysRevD.79.113009;%%
  %7 citations counted in INSPIRE as of 19 Jun 2016
  
%\cite{Hirsch:1997vz}
\bibitem{Hirsch:1997vz} 
  M.~Hirsch, H.~V.~Klapdor-Kleingrothaus and S.~G.~Kovalenko,
  %``B-L violating masses in softly broken supersymmetry,''
  Phys.\ Lett.\ B {\bf 398}, 311 (1997)
  doi:10.1016/S0370-2693(97)00234-7
  [hep-ph/9701253].
  %%CITATION = doi:10.1016/S0370-2693(97)00234-7;%%
  %136 citations counted in INSPIRE as of 13 Jan 2016

%\cite{Chun:2004mu}
\bibitem{Chun:2004mu}
  E.~J.~Chun and S.~C.~Park,
  %``Neutrino mass from R-parity violation in split supersymmetry,''
  JHEP {\bf 0501} (2005) 009
  doi:10.1088/1126-6708/2005/01/009
  [hep-ph/0410242].
  %%CITATION = doi:10.1088/1126-6708/2005/01/009;%%
  %30 citations counted in INSPIRE as of 24 Aug 2016
  
%\cite{Dhuria:2011ye}
\bibitem{Dhuria:2011ye}
  M.~Dhuria and A.~Misra,
  %``Towards Large Volume Big Divisor D3-D7 'mu-Split Supersymmetry' and Ricci-Flat Swiss-Cheese Metrics, and Dimension-Six Neutrino Mass Operators,''
  Nucl.\ Phys.\ B {\bf 855} (2012) 439
  doi:10.1016/j.nuclphysb.2011.10.006
  [arXiv:1106.5359 [hep-th]].
  %%CITATION = doi:10.1016/j.nuclphysb.2011.10.006;%%
  %5 citations counted in INSPIRE as of 24 Aug 2016
  
%\cite{Gunion:1984yn}
\bibitem{Gunion:1984yn}
  J.~F.~Gunion and H.~E.~Haber,
  %``Higgs Bosons in Supersymmetric Models. 1.,''
  Nucl.\ Phys.\ B {\bf 272} (1986) 1
   Erratum: [Nucl.\ Phys.\ B {\bf 402} (1993) 567].
  doi:10.1016/0550-3213(86)90340-8, 10.1016/0550-3213(93)90653-7
  %%CITATION = doi:10.1016/0550-3213(86)90340-8, 10.1016/0550-3213(93)90653-7;%%
  %1304 citations counted in INSPIRE as of 19 Jun 2016
%\cite{Gunion:1986nh}
%\bibitem{Gunion:1986nh}
  J.~F.~Gunion and H.~E.~Haber,
  %``Higgs Bosons in Supersymmetric Models. 2. Implications for Phenomenology,''
  Nucl.\ Phys.\ B {\bf 278} (1986) 449
   Erratum: [Nucl.\ Phys.\ B {\bf 402} (1993) 569].
  doi:10.1016/0550-3213(93)90654-8, 10.1016/0550-3213(86)90050-7
  %%CITATION = doi:10.1016/0550-3213(93)90654-8, 10.1016/0550-3213(86)90050-7;%%
  %397 citations counted in INSPIRE as of 19 Jun 2016
%\cite{Gunion:1988yc}
%\bibitem{Gunion:1988yc}
  J.~F.~Gunion and H.~E.~Haber,
  %``Higgs Bosons in Supersymmetric Models. 3. Decays Into Neutralinos and Charginos,''
  Nucl.\ Phys.\ B {\bf 307} (1988) 445
   Erratum: [Nucl.\ Phys.\ B {\bf 402} (1993) 569].
  doi:10.1016/0550-3213(93)90655-9, 10.1016/0550-3213(88)90259-3
  %%CITATION = doi:10.1016/0550-3213(93)90655-9, 10.1016/0550-3213(88)90259-3;%%
  %125 citations counted in INSPIRE as of 19 Jun 2016
  
%\cite{Gunion:1989we}
\bibitem{Gunion:1989we}
  J.~F.~Gunion, H.~E.~Haber, G.~L.~Kane and S.~Dawson,
  %``The Higgs Hunter's Guide,''
  Front.\ Phys.\  {\bf 80} (2000) 1.
  %%CITATION = FRPHA,80,1;%%
  %560 citations counted in INSPIRE as of 19 Jun 2016
  
%\cite{Hirsch:2000ef}
\bibitem{Hirsch:2000ef} 
  M.~Hirsch, M.~A.~Diaz, W.~Porod, J.~C.~Romao and J.~W.~F.~Valle,
  %``Neutrino masses and mixings from supersymmetry with bilinear R parity violation: A Theory for solar and atmospheric neutrino oscillations,''
  Phys.\ Rev.\ D {\bf 62}, 113008 (2000)
  [Phys.\ Rev.\ D {\bf 65}, 119901 (2002)]
  doi:10.1103/PhysRevD.62.113008, 10.1103/PhysRevD.65.119901
  [hep-ph/0004115].
  %%CITATION = doi:10.1103/PhysRevD.62.113008, 10.1103/PhysRevD.65.119901;%%
  %333 citations counted in INSPIRE as of 13 Jan 2016
  
%\cite{Diaz:2003as}
\bibitem{Diaz:2003as}
  M.~A.~Diaz, M.~Hirsch, W.~Porod, J.~C.~Romao and J.~W.~F.~Valle,
  %``Solar neutrino masses and mixing from bilinear R parity broken supersymmetry: Analytical versus numerical results,''
  Phys.\ Rev.\ D {\bf 68} (2003) 013009
   Erratum: [Phys.\ Rev.\ D {\bf 71} (2005) 059904]
  doi:10.1103/PhysRevD.71.059904, 10.1103/PhysRevD.68.013009
  [hep-ph/0302021].
  %%CITATION = doi:10.1103/PhysRevD.71.059904, 10.1103/PhysRevD.68.013009;%%
  %148 citations counted in INSPIRE as of 19 Jun 2016
  
%\cite{Diaz:2014jta}
\bibitem{Diaz:2014jta}
  M.~A.~D\'iaz, M.~Rivera and N.~Rojas,
  %``On Neutrino Masses in the MSSM with BRpV,''
  Nucl.\ Phys.\ B {\bf 887} (2014) 338
  doi:10.1016/j.nuclphysb.2014.08.012
  [arXiv:1401.7357 [hep-ph]].
  %%CITATION = doi:10.1016/j.nuclphysb.2014.08.012;%%

%\cite{Buchalla:1995vs}
\bibitem{Buchalla:1995vs}
  For a pedagogical introduction in effective field theories and the treatment
  of loop corrections in this context, see for instance: 
  G.~Buchalla, A.~J.~Buras and M.~E.~Lautenbacher,
  %``Weak decays beyond leading logarithms,''
  Rev.\ Mod.\ Phys.\  {\bf 68} (1996) 1125
  doi:10.1103/RevModPhys.68.1125
  [hep-ph/9512380].
  %%CITATION = doi:10.1103/RevModPhys.68.1125;%%
  %2041 citations counted in INSPIRE as of 20 Jun 2016
  An application with very similar characteristics to our case can be found at:
%\cite{Bouchand:2012dx}
%\bibitem{Bouchand:2012dx}
  R.~Bouchand and A.~Merle,
  %``Running of Radiative Neutrino Masses: The Scotogenic Model,''
  JHEP {\bf 1207} (2012) 084
  doi:10.1007/JHEP07(2012)084
  [arXiv:1205.0008 [hep-ph]].
  %%CITATION = doi:10.1007/JHEP07(2012)084;%%
  %48 citations counted in INSPIRE as of 20 Jun 2016  
  
%\cite{Sierra:2016rcz}
\bibitem{Sierra:2016rcz}
  D.~Aristizabal Sierra, C.~Simoes and D.~Wegman,
  %``Radiative accidental matter,''
  JHEP {\bf 1607} (2016) 124
  doi:10.1007/JHEP07(2016)124
  [arXiv:1605.08267 [hep-ph]].
  %%CITATION = doi:10.1007/JHEP07(2016)124;%% 

%\cite{Aad:2015zhl}
\bibitem{Aad:2015zhl}
  G.~Aad {\it et al.} [ATLAS and CMS Collaborations],
  %``Combined Measurement of the Higgs Boson Mass in $pp$ Collisions at $\sqrt{s}=7$ and 8 TeV with the ATLAS and CMS Experiments,''
  Phys.\ Rev.\ Lett.\  {\bf 114} (2015) 191803
  doi:10.1103/PhysRevLett.114.191803
  [arXiv:1503.07589 [hep-ex]].
  %%CITATION = doi:10.1103/PhysRevLett.114.191803;%%
  %492 citations counted in INSPIRE as of 23 Aug 2016
  
%\cite{Aaboud:2016zpr}
\bibitem{Aaboud:2016zpr}
  M.~Aaboud {\it et al.} [ATLAS Collaboration],
  %``Search for squarks and gluinos in events with hadronically decaying tau leptons, jets and missing transverse momentum in proton-proton collisions at $\sqrt{s}=13$ TeV recorded with the ATLAS detector,''
  arXiv:1607.05979 [hep-ex].
  %%CITATION = ARXIV:1607.05979;%%
  %1 citations counted in INSPIRE as of 23 Aug 2016
%\cite{Khachatryan:2011tk}
%\bibitem{Khachatryan:2011tk}
  V.~Khachatryan {\it et al.} [CMS Collaboration],
  %``Search for Supersymmetry in pp Collisions at 7 TeV in Events with Jets and Missing Transverse Energy,''
  Phys.\ Lett.\ B {\bf 698} (2011) 196
  doi:10.1016/j.physletb.2011.03.021
  [arXiv:1101.1628 [hep-ex]].
  %%CITATION = doi:10.1016/j.physletb.2011.03.021;%%
  %258 citations counted in INSPIRE as of 23 Aug 2016
  
%\cite{Haber:1984rc}
%\bibitem{Haber:1984rc}
%  H.~E.~Haber and G.~L.~Kane,
%  %``The Search for Supersymmetry: Probing Physics Beyond the Standard Model,''
%  Phys.\ Rept.\  {\bf 117} (1985) 75.
%  doi:10.1016/0370-1573(85)90051-1
%  %%CITATION = doi:10.1016/0370-1573(85)90051-1;%%
%  %4532 citations counted in INSPIRE as of 19 Jun 2016
 
%\cite{:2012gk}
%\bibitem{:2012gk} 
%  G.~Aad {\it et al.}  [ATLAS Collaboration],
%  %``Observation of a new particle in the search for the Standard Model Higgs boson with the ATLAS detector at the LHC,''
%  Phys.\ Lett.\ B {\bf 716}, 1 (2012)
%  [arXiv:1207.7214 [hep-ex]].
%  %%CITATION = ARXIV:1207.7214;%%

%%--- TO BE CITED
 
\end{thebibliography}
\end{document}